\documentclass[conference]{IEEEtran}
\usepackage[utf8]{inputenc}
\usepackage[T1]{fontenc}
\usepackage{braket}
\usepackage{amsfonts, amsmath, amssymb, mathtools}
\usepackage{dcolumn}
\usepackage{booktabs}
\usepackage{siunitx}
\usepackage{algorithm}
\usepackage{algpseudocode}
\usepackage{xurl} 
\usepackage[breaklinks=true,colorlinks,citecolor=blue,linkcolor=blue,urlcolor=blue]{hyperref}
\usepackage{orcidlink}
\usepackage{xcolor}
\usepackage{graphicx}
\usepackage{float}
\usepackage{stfloats}
\usepackage{comment}
\usepackage{microtype}

\begin{document}

\title{Quantum Feature Selection with Higher-Order Binary Optimization on Trapped-Ion Hardware}

\author{

\IEEEauthorblockN{
Carlos Flores-Garrig\'os$^{1,2}$,
Anton Simen$^{1,3}$,
Qi Zhang$^{1}$,
Enrique Solano$^{1}$,
Narendra N. Hegade$^{1,2}$
}\medskip
\IEEEauthorblockA{
$^{1}$Kipu Quantum GmbH, Greifswalderstrasse 212, 10405 Berlin, Germany\\
$^{2}$IDAL, Electronic Engineering Department, ETSE-UV, University of Valencia,\\ Avgda. Universitat s/n, 46100 Burjassot, Valencia, Spain\\
$^{3}$Department of Physical Chemistry, University of the Basque Country UPV/EHU,\\ Apartado 644, 48080 Bilbao, Spain
}\\

\IEEEauthorblockN{
Sayonee Ray$^{5}$,
Claudio Girotto$^{5}$,
Jason Iaconis$^{5}$,
Martin Roetteler$^{5}$
}\medskip
\IEEEauthorblockA{
$^{5}$IonQ Inc., 4505 Campus Dr, College Park, MD 20740, USA}
}

\maketitle

\begin{abstract}
We present a quantum feature-selection framework based on a higher-order unconstrained binary optimization (HUBO) formulation that explicitly incorporates multivariate dependencies beyond standard quadratic encodings. In contrast to QUBO-based approaches, the proposed model includes one-, two-, and three-body interaction terms derived from mutual-information measures, enabling the objective function to capture feature relevance, pairwise redundancy, and higher-order statistical structure within a unified energy model. To suppress trivial all-selected solutions, we further include structured linear penalties that promote sparsity while preserving informative variables. The resulting HUBO instances are optimized with digitized counterdiabatic quantum optimization on IonQ Forte and compared against noiseless quantum simulation as well as two classical dimensionality-reduction baselines: SelectKBest based on mutual information and principal component analysis (PCA). We evaluate the proposed workflow on two benchmark classification datasets, namely the Gallstone dataset and the Spambase dataset, and analyze both predictive performance and selected-subset structure. The results show good qualitative agreement between hardware executions and noiseless simulations, supporting the feasibility of implementing higher-order feature-selection Hamiltonians on current trapped-ion processors. In addition, the quantum approach yields competitive classification performance while producing compact and informative feature subsets, highlighting the potential of higher-order quantum optimization for machine-learning preprocessing tasks.
\end{abstract}

\section{Introduction}

Feature selection (FS) is a central problem in modern artificial intelligence (AI), with critical applications spanning machine learning, computer vision, natural language processing, and biomedical data analysis~\cite{guyon2003introduction, dash1997feature, bolon2013review, li2017feature}. 
By identifying the subset of features that contributes most to a model's predictive performance, FS enhances interpretability, reduces overfitting, and improves computational efficiency. 
In high-dimensional datasets—common in genomics~\cite{tadist2019feature}, finance~\cite{liang2015effect}, and medicine~\cite{remeseiro2019review}—irrelevant or redundant variables can obscure the underlying signal and degrade generalization performance. 
Effective feature selection is therefore essential not only for building compact and interpretable models but also for ensuring stable performance across tasks and domains.

Classical FS techniques are typically model-dependent, iterative, and computationally demanding. Wrapper methods~\cite{chandrashekar2014survey} repeatedly train a learning model to evaluate subsets of features, which becomes prohibitive as dimensionality increases. Filter-based methods~\cite{porkodi2014comparison}, such as mutual information~\cite{vergara2014review} or correlation-based criteria~\cite{hall1999correlation}, provide faster alternatives but often disregard feature interactions. Embedded methods, such as LASSO~\cite{muthukrishnan2016lasso} or tree-based approaches~\cite{zhou2021feature}, intertwine feature selection and model training but remain sensitive to model hyperparameters and bias toward linear or hierarchical dependencies. State-of-the-art approaches combine relevance and redundancy measures but still require repeated training cycles or heuristic scoring, limiting their scalability and robustness in complex and high-dimensional feature spaces.

Quantum computing offers a fundamentally different route to optimization by exploiting quantum superposition and entanglement to explore exponentially large configuration spaces. 
Several recent studies have formulated combinatorial optimization problems, including feature selection, as Quadratic Unconstrained Binary Optimization (QUBO) problems~\cite{mucke2023feature, ferrari2022towards, orquinmarques2025analogquantumfeatureselection}. In such formulations, each feature corresponds to a binary variable $x_i \in \{0,1\}$, and the optimization objective is represented by a quadratic Hamiltonian whose ground state encodes the optimal feature subset. 
Quantum annealers and variational quantum algorithms can then approximate this ground state by evolving the system under the corresponding Hamiltonian. 
Standard QUBO formulations are restricted to explicit one- and two-body terms. They naturally encode individual feature relevance and pairwise
redundancy, but higher-order dependencies must either be neglected, approximated, or introduced through additional variables and quadratization
overheads. This can be restrictive in settings where groups of features carry information that is not visible from individual or pairwise scores alone. Such multivariate dependencies are common in real-world data but remain inaccessible to quadratic models.

To address this limitation, we formulate feature selection directly as a Higher-Order Unconstrained Binary Optimization (HUBO) problem. The resulting Hamiltonian contains explicit three-body terms, allowing the cost function to represent higher-order dependencies among features without first reducing the
objective to a quadratic form. Although this increases the complexity of the optimization landscape, it also provides a more expressive energy model for
balancing relevance, redundancy, and multivariate feature structure. We solve the resulting HUBO instances using DCQO on trapped-ion hardware and compare the selected feature subsets against noiseless quantum simulation and classical
reference methods.

The remainder of this paper is organized as follows. Section~\ref{sec:methodology} introduces the theoretical background and notation for Ising variables and the HUBO formulation of feature selection and details the computation of mutual-information-based coefficients. Section~\ref{sec:hardware} describes the IonQ hardware implementation, together with the optimization procedure and the low-energy post-selection rule for identifying stable features. Section~\ref{sec:experiments} presents experimental results on two benchmark datasets, namely the Gallstone and Spambase datasets, and compares the proposed approach against classical baselines including SelectKBest based on mutual information and principal component analysis (PCA). Finally, Section~\ref{sec:conclusion} discusses implications, limitations, and future directions toward scalable quantum feature selection frameworks.

\section{Methodology}
\label{sec:methodology}

\subsection{Ising and HUBO Formulation}

We formulate feature selection as a Higher-Order Unconstrained Binary Optimization (HUBO) problem in the Ising convention, where each feature is associated with a spin variable $Z_i \in \{-1,+1\}$. We adopt the convention $Z_i=-1$ if feature $i$ is selected and $Z_i=+1$ otherwise. Equivalently, in binary notation,
\begin{equation}
x_i=\frac{1-Z_i}{2}\in\{0,1\},
\end{equation}
so that $x_i=1$ denotes inclusion and $x_i=0$ exclusion.

The feature-selection problem is encoded in the Hamiltonian
\begin{equation}
H(Z)=\sum_i h_i Z_i+\sum_{i<j} J_{ij} Z_i Z_j+\sum_{i<j<k} K_{ijk} Z_i Z_j Z_k+\mathrm{C},
\label{eq:ising_hubo}
\end{equation}
where $h_i$, $J_{ij}$, and $K_{ijk}$ are the one-, two-, and three-body coefficients, respectively, and $\mathrm{C}$ is a constant offset. The one-body terms encode individual feature relevance, the two-body terms capture pairwise redundancy, and the three-body terms account for higher-order dependencies beyond quadratic encodings.

\begin{figure*}[!t]
    \centering
    \includegraphics[width=0.8\linewidth]{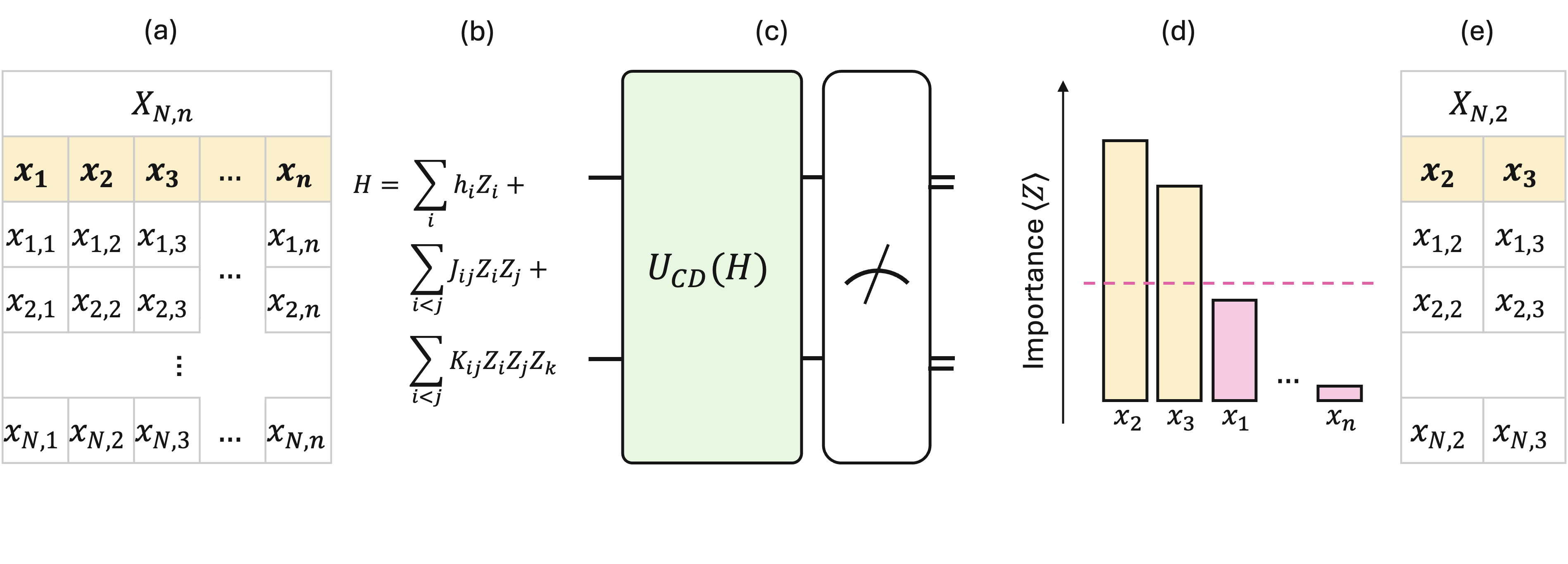}
    \caption{Schematics of the quantum feature selection procedure. (a) illustrates a training dataset with $n$ variables and $N$ samples used to compute (b) the coefficients of a higher-order spin Hamiltonian---based on classical statistics from the dataset---whose low-energy configurations encode solutions of the feature selection problem. A (c) digitized counterdiabatic (CD) protocol is implemented in order to obtain a distribution of low-energy states, and (d) local magnetizations that capture the importance of the features are calculated. Based on a importance threshold, (e) features are selected.}
    \label{fig:schematics}
\end{figure*}

With this convention, positive $h_i$ favors selecting informative features since $h_i Z_i=-h_i$ when $Z_i=-1$. Positive $J_{ij}$ discourages the simultaneous selection of strongly correlated features, while negative $K_{ijk}$ penalizes configurations in which all three features are selected simultaneously, helping prevent trivial dense solutions. The ground state of Eq.~(\ref{eq:ising_hubo}) therefore defines the feature subset that optimally balances relevance, redundancy, and higher-order structure. An illustration of the procedure can be seen in Fig. \ref{fig:schematics}

\subsection{Mutual-Information-Derived Coefficients}

The coefficients $(h_i,J_{ij},K_{ijk})$ are constructed from mutual-information (MI) quantities that quantify both feature-target relevance and statistical dependencies among features. To place all interaction orders on a common scale, the values $\mathrm{MI}(X_i,y)$, $\mathrm{MI}(X_i,X_j)$, and $\mathrm{MI}([X_i,X_j],X_k)$ are jointly normalized using a single global min--max transformation.

The one-body coefficients are defined as
\begin{equation}
h_i=w_1\,
\frac{\mathrm{MI}(X_i,y)-\min(\mathrm{MI})}
{\max(\mathrm{MI})-\min(\mathrm{MI})},
\label{eq:mi_onebody}
\end{equation}
where $w_1>0$ controls the contribution of feature relevance.

The two-body coefficients are defined as
\begin{equation}
J_{ij}=w_2\,
\frac{\mathrm{MI}(X_i,X_j)-\min(\mathrm{MI})}
{\max(\mathrm{MI})-\min(\mathrm{MI})}, \qquad i<j,
\label{eq:mi_twobody}
\end{equation}
where $w_2>0$ sets the strength of pairwise redundancy suppression.

The three-body coefficients are defined as
\begin{equation}
K_{ijk}=-\,w_3\,
\frac{\overline{\mathrm{MI}}_{\mathrm{cyclic}}([X_i,X_j],X_k)-\min(\mathrm{MI})}
{\max(\mathrm{MI})-\min(\mathrm{MI})}, \qquad i<j<k,
\label{eq:mi_threebody}
\end{equation}
where $w_3>0$ controls the higher-order contribution and
$\overline{\mathrm{MI}}_{\mathrm{cyclic}}$ denotes the cyclic average of
$\mathrm{MI}([X_i,X_j],X_k)$,
$\mathrm{MI}([X_i,X_k],X_j)$, and
$\mathrm{MI}([X_j,X_k],X_i)$.

This construction yields a unified Hamiltonian in which individually relevant features lower the energy, pairwise dependencies discourage redundant selections, and three-body terms introduce higher-order feature-structure
information beyond quadratic encodings.

\subsection{Structured Penalties to Prevent Trivial Solutions}

To discourage trivial minima, such as selecting all features or retaining variables with negligible relevance, we introduce an additional linear penalty on the one-body coefficients,
\begin{equation}
H_{\lambda}=\sum_i \Delta h_i Z_i,
\qquad
\Delta h_i=
\begin{cases}
-\lambda \left(\frac{\tau-c_i}{\tau}\right)^p, & c_i<\tau,\\
0, & c_i\ge \tau,
\end{cases}
\label{eq:lambda_penalty}
\end{equation}
where $c_i$ is the normalized mutual information of feature $i$, $\tau$ is a relevance threshold, $\lambda$ sets the maximum penalty amplitude, and $p$ is the hinge exponent, taken as $p=2$ in this work.

The total Hamiltonian is given by
\begin{equation}
H_{\mathrm{tot}}(Z)=H(Z)+H_{\lambda}(Z),
\end{equation}
which defines the objective function optimized in both hardware and noisy-simulation experiments.

\section{Quantum Optimization Framework}
\label{sec:hardware}

\subsection{Digitized Counterdiabatic Quantum Optimization}

We use digitized counterdiabatic quantum optimization (DCQO) to sample
low-energy states of the HUBO Hamiltonian~\cite{hegade2022dcqo}. DCQO is a
gate-based approach inspired by shortcuts to adiabaticity: a digitized
interpolation connects an initial driver Hamiltonian to the target Hamiltonian,
while approximate counterdiabatic terms are added to reduce diabatic transitions
during finite-time evolution.

In our experiments, DCQO is applied to the full feature-selection Hamiltonian
$H_{\mathrm{tot}}$. The resulting measurement samples are ranked by their
energies, and the lowest-energy fraction is used to estimate the feature
importance scores defined in Eq.~(\ref{eq:feature_importance}). The selected
feature subset is then obtained by thresholding these scores.

For larger instances, the same workflow could be combined with bias-field DCQO, where low-energy samples from previous iterations update longitudinal
bias fields in the initial Hamiltonian~\cite{cadavid2025bfdcqo}. This strategy
has recently been applied to higher-order binary optimization
problems~\cite{romero2025bfhubo}, making it a natural extension of the present
HUBO-based feature-selection approach.

\subsection{IonQ Forte Hardware}
Our key results were obtained from IonQ's Forte class systems, based on Yb+ ions, where the ions are produced via laser ablation and selective ionization before being secured within compact vacuum packages using integrated surface linear Paul traps. For the Forte class systems, two-photon Raman transitions with 355 nm laser pulses are used to achieve universal gate control via arbitrary single-qubit rotations and entangling ZZ gates. The optical architecture of these QPUs relies on acousto-optic deflectors (AODs), which allow for independent and precise beam steering to individual ions, significantly mitigating alignment errors (Ref.~\cite{Kim-aod,PRXQuantum.2.020343}). These technical advancements ensure the high gate fidelities necessary for successfully running deep quantum circuits on the hardware (Refs.~\cite{Chen2024benchmarkingtrapped}). All the final circuits implemented on the QPU used $32$ qubits (from the $32$ input features) and approximately $1,000$ two-qubit gates for each dataset, with $2,000$ shots per circuit. Note that the results after $1$ iteration is already competitive with the classical benchmarks, as shown in Table.~\ref{tab:gallstone_results} and~\ref{tab:spam_results}.

\subsection{Measurement and Feature Selection Criterion}

After quantum evolution, the system is measured in the computational basis, yielding a set of $S$ bitstring samples $\{Z^{(s)}\}_{s=1}^S$, with $Z_i^{(s)} \in \{-1,+1\}$. Each sample is mapped to binary selection variables through
\begin{equation}
x_i^{(s)}=\frac{1-Z_i^{(s)}}{2}\in\{0,1\},
\end{equation}
so that $x_i^{(s)}=1$ indicates that feature $i$ is selected in sample $s$.

For each measured bitstring, we evaluate the corresponding Hamiltonian value $H_{\mathrm{tot}}(Z^{(s)})$ and rank all samples according to energy. We then retain a subset of the lowest-energy samples, denoted by $\mathcal{S}_{\rho}$, where $\rho \in (0,1]$ represents the fraction of samples kept after energy-based filtering. This can be seen in Fig. \ref{fig:schematics}, sections (c) and (d).

The feature importance score is defined as the empirical average over the retained subset,
\begin{equation}
I_i = \frac{1}{|\mathcal{S}_{\rho}|} \sum_{s \in \mathcal{S}_{\rho}} x_i^{(s)},
\label{eq:feature_importance}
\end{equation}
where $|\mathcal{S}_{\rho}| = \rho S$. By construction, $I_i \in [0,1]$, with $I_i=0$ indicating that feature $i$ is never selected within the retained low-energy subset, and $I_i=1$ indicating that it is always selected.

The final feature subset is obtained by thresholding these importance scores:
\begin{equation}
I_i \ge \delta,
\end{equation}
where $\delta \in [0,1]$ is a selection threshold. Smaller values of $\delta$ yield larger subsets, while larger values retain only the most consistently selected features.

This post-processing procedure is therefore governed by two interpretable hyperparameters: the retention fraction $\rho$, which controls the number of low-energy samples contributing to the ranking, and the threshold $\delta$, which determines the strictness of the final selection. As illustrated in Fig.~\ref{fig:feature_performance} and Fig.~\ref{fig:spam_forte_pipeline}, this approach produces feature-importance scores in the range $[0,1]$ for both Gallstone and Spam Detection datasets.

\section{Experimental Protocol}
\label{sec:experiments}

\subsection{Datasets}

To evaluate the proposed HUBO-based quantum feature selection framework, we conduct experiments on two benchmark classification datasets spanning biomedical and textual domains.

The first dataset is the Gallstone UCI dataset~\cite{esen2024early}, which comprises 319 clinical samples described by 38 heterogeneous attributes, including demographic information, bioimpedance indicators, and biochemical measurements. The prediction task is binary, aiming to classify whether a patient presents gallstones based on these clinical variables. This dataset combines continuous and categorical features and represents a small-sample, mixed-type medical scenario.

The second dataset is the Spambase dataset~\cite{spambase_94}, which contains 4,601 email instances described by 57 real-valued features. These features include word and character frequency statistics, as well as measures of capital letter usage and other structural properties of emails. The task is to distinguish spam from legitimate messages. The dataset exhibits moderate dimensionality and significant redundancy among lexical features, making it well suited for evaluating relevance and redundancy-aware feature selection methods.

All datasets are standardized prior to mutual-information computation, and categorical attributes are encoded using one-hot representations when applicable. Due to the 32-qubit hardware limit, each dataset was preselected to 32 features using the $MI(X_i,y)$ ranking. Together, these datasets provide complementary evaluation settings in terms of sample size, feature heterogeneity, and redundancy structure, allowing us to assess the robustness of the proposed HUBO-based approach across different data regimes.

For comparison, we include two classical reference methods: SelectKBest based on mutual information and principal component analysis (PCA).

\section{Results}\label{sec:results}
\subsection{Overall Performance on the Gallstone Dataset}

\begin{figure*}[!t]
    \centering
    \includegraphics[width=\textwidth]{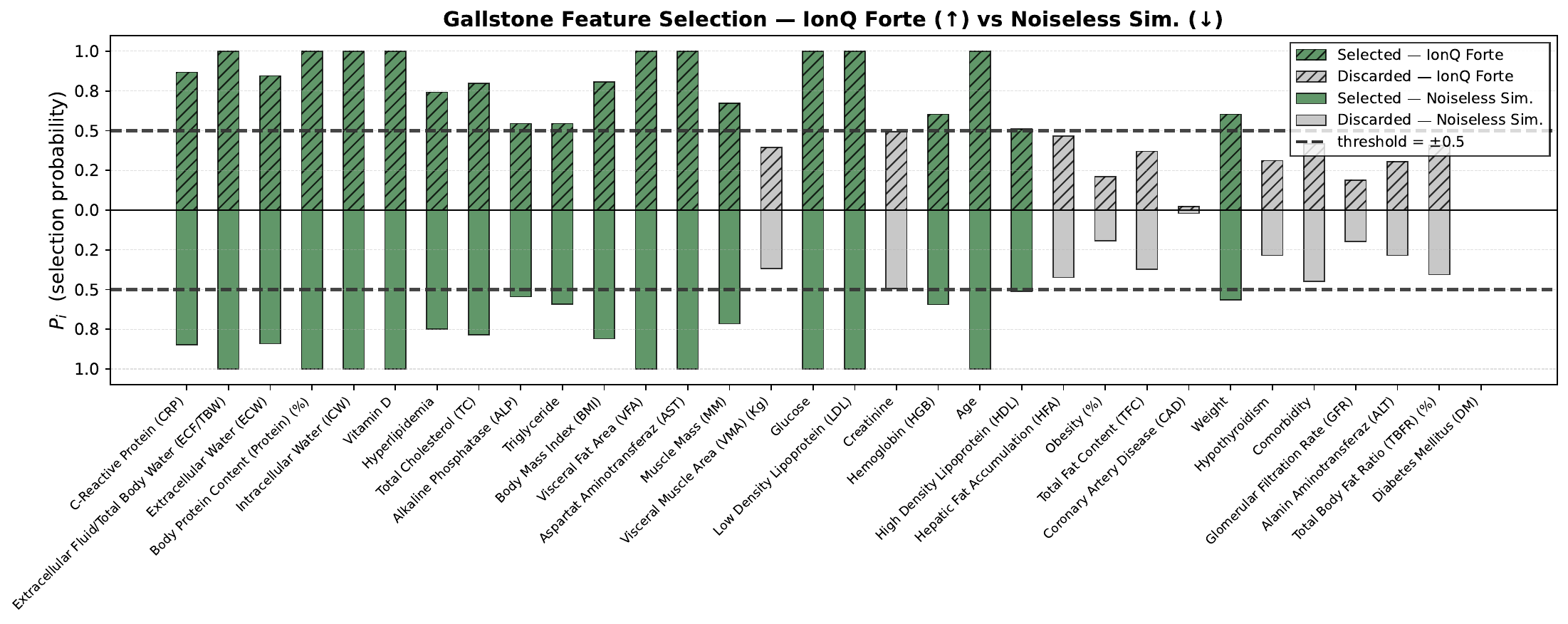}
    \caption{Feature inclusion probabilities $P_i \in [0,1]$ for the Gallstone dataset obtained from IonQ Forte hardware (upper half) and noiseless simulation (lower half). Each bar represents the empirical selection frequency of a feature within the retained low-energy subset. Features are classified as selected or discarded according to a threshold (dashed lines), illustrating the consistency between hardware and simulation in identifying relevant variables.}
    \label{fig:feature_performance}
\end{figure*}

We first analyze the consistency between the IonQ Forte hardware and the noiseless simulation. As shown in Fig.~\ref{fig:feature_performance}, both implementations produce highly similar feature inclusion probability profiles across all variables. In particular, the majority of features are consistently classified as either selected or discarded under the same thresholding rule, indicating a strong qualitative agreement between hardware and noiseless simulation. Nevertheless, small discrepancies can be observed in a subset of features near the decision boundary, where minor variations in inclusion probabilities lead to different selection outcomes.

\begin{figure}[H]
    \centering
    \includegraphics[width=\linewidth]{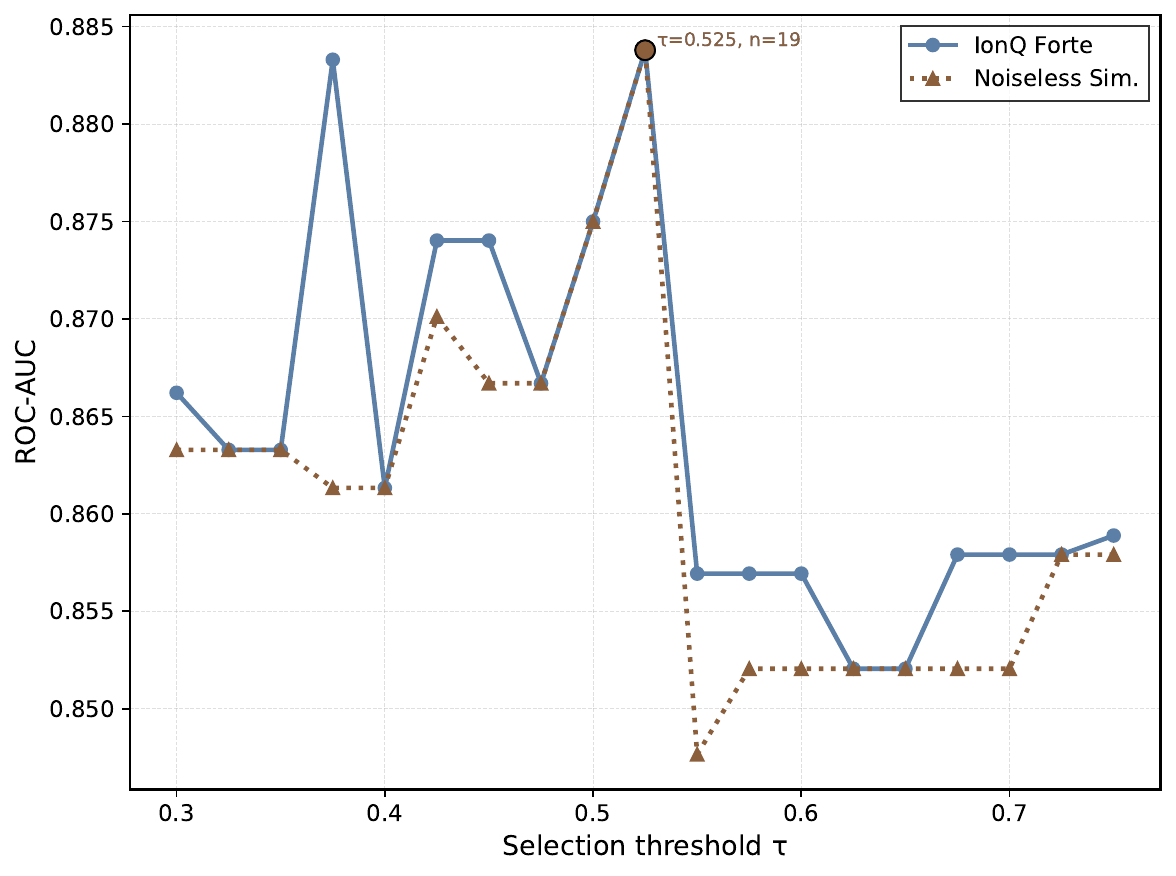}
    \caption{ROC-AUC performance on the Gallstone dataset (results in test) as a function of the selection threshold $\tau$ applied to the feature inclusion probabilities. Results are shown for IonQ Forte hardware and noiseless simulation. The optimal threshold balances predictive performance and subset size, with annotations indicating the number of selected features at the best-performing configurations.}
    \label{fig:performance}
\end{figure}

Although in the best-case scenario both approaches yield identical feature subsets (see Table~\ref{tab:gallstone_results}), a more detailed analysis reveals subtle differences when varying the selection threshold. To further investigate this effect, we perform a systematic sweep over the selection threshold $\tau$, evaluating the predictive performance of the resulting feature subsets using a fixed Random Forest classifier with hyperparameters $\texttt{n\_estimators}=300$ and $\texttt{random\_state}=42$. As shown in Fig.~\ref{fig:performance}, the optimal performance is achieved at $\tau=0.525$, where the IonQ Forte hardware attains a higher ROC-AUC than the noiseless simulation. This indicates that small differences in the sampled feature rankings can affect the downstream classifier performance, especially for features close to the selection threshold.

Finally, we compare the proposed quantum feature selection approach against classical baselines. As shown in Table~\ref{tab:gallstone_results}, the IonQ Forte solution (reducing dimensionality from 32 to 19 features) outperforms both PCA (with $95\%$ explained variance) and the SelectKBest method based on mutual information. Notably, the latter corresponds to a univariate ranking strategy that is effectively equivalent to considering only the one-body contribution of the Hamiltonian. The observed performance gap therefore highlights the benefit of incorporating higher-order interactions in the feature selection process, as enabled by the HUBO formulation.

\begin{table}[H]
\centering
\caption{Performance comparison on the Gallstone dataset. 
Metrics reported are accuracy (Acc), F1-score (F1), and ROC-AUC (AUC).}
\label{tab:gallstone_results}
\scriptsize
\resizebox{0.996\linewidth}{!}{%
\setlength{\tabcolsep}{4pt}
\renewcommand{\arraystretch}{1.15}
\begin{tabular}{lccc}
\toprule
\textbf{Method} & \textbf{Acc} & \textbf{F1} & \textbf{AUC} \\
\midrule
IonQ Forte (n = 19, $\tau=0.525$) & \textbf{0.79} & \textbf{0.79} & \textbf{0.88} \\
IonQ Noiseless Sim. (n = 19, $\tau=0.525$) & \textbf{0.79} & \textbf{0.79} & \textbf{0.88} \\
All features & 0.78 & 0.78 & 0.86 \\
PCA (var = 0.95) & 0.72 & 0.72 & 0.79 \\
SelectKBest (n = 19) & 0.75 & 0.75 & 0.84 \\
\bottomrule
\end{tabular}
}
\end{table}

\subsection{Overal Performance on the Spambase Dataset}

For the Spambase dataset, we directly report results obtained from IonQ Forte hardware. This allows us to assess the effectiveness of the proposed HUBO-based feature selection pipeline under realistic experimental conditions.

\begin{figure*}[!t]
    \centering
    \includegraphics[width=\textwidth]{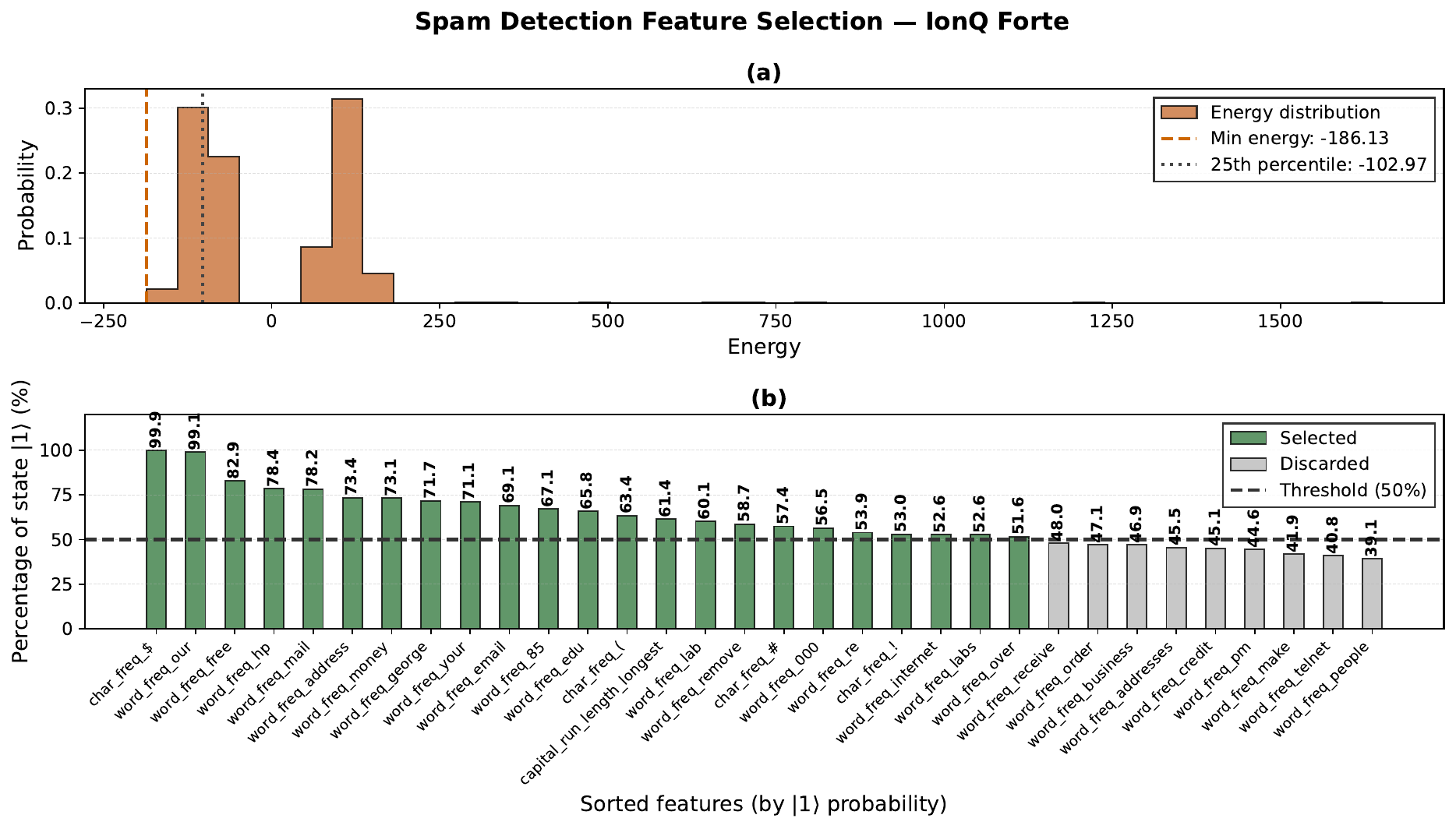}
    \caption{Feature selection pipeline for the Spambase dataset using IonQ Forte. 
    (a) Distribution of sampled Hamiltonian values, where only the lowest $\rho=25\%$ of states, i.e., those below the 25th percentile, are retained for post-processing. 
    (b) Feature inclusion probabilities $P_i$ computed as the empirical frequency of $x_i=1$ within the retained low-energy subset. 
    Features are sorted by $P_i$ and classified as selected or discarded using a threshold $\tau=0.50$.}
    \label{fig:spam_forte_pipeline}
\end{figure*}

The feature selection procedure is illustrated in Fig.~\ref{fig:spam_forte_pipeline}. Starting from the full set of measured bitstring samples, we first evaluate the corresponding Hamiltonian values and retain only the lowest-energy fraction of states. In particular, we select the subset corresponding to the lowest $\rho=25\%$ of sampled energies, i.e., those below the 25th percentile of the energy distribution. This filtering step focuses the analysis on the most relevant candidate solutions while reducing the impact of higher-energy configurations.

Feature inclusion probabilities $P_i$ are then computed as the empirical frequency of selecting feature $i$ within this low-energy subset. As shown in Fig.~\ref{fig:spam_forte_pipeline}(b), this results in a clear ranking of features according to their selection stability. A threshold $\tau=0.50$ is subsequently applied to obtain the final feature subset.

The resulting model achieves strong predictive performance, as reported in Table~\ref{tab:spam_results}. In particular, the quantum-selected subset reduces the dimensionality from 32 to 23 features while maintaining—and slightly improving—classification performance compared to using all features. This demonstrates that the proposed approach is able to identify a compact and informative subset without sacrificing predictive power.

Furthermore, the HUBO-based selection outperforms classical baselines, including PCA (with $95\%$ explained variance) and SelectKBest based on mutual information. The latter relies solely on univariate feature relevance, effectively corresponding to the one-body contribution of the Hamiltonian. The observed improvement, therefore, highlights the advantage of incorporating higher-order interactions, which enable the identification of more informative feature combinations beyond purely individual relevance.

\begin{table}[H]
\centering
\caption{Performance comparison on the Spambase dataset. 
Metrics reported are accuracy (Acc), F1-score (F1), and ROC-AUC (AUC).}
\label{tab:spam_results}
\scriptsize
\resizebox{0.996\linewidth}{!}{%
\setlength{\tabcolsep}{4pt}
\renewcommand{\arraystretch}{1.15}
\begin{tabular}{lccc}
\toprule
\textbf{Method} & \textbf{Acc} & \textbf{F1} & \textbf{AUC} \\
\midrule
IonQ Forte (n = 23, $\tau=0.50$) & \textbf{0.9475} & \textbf{0.9464} & \textbf{0.9836} \\
IonQ Noiseless Sim. (n = 23, $\tau=0.525$) & 0.9446 & 0.9445 & 0.9819 \\
All features & 0.9457 & 0.9456 & 0.9817 \\
PCA (var = 0.95) & 0.9186 & 0.9185 & 0.9615 \\
SelectKBest (n = 23) & 0.9414 & 0.9413 & 0.9805 \\
\bottomrule
\end{tabular}
}
\end{table}

\section{Conclusion}\label{sec:conclusion}

We have presented a quantum feature selection framework based on a Higher-Order Unconstrained Binary Optimization (HUBO) formulation, where feature relevance, redundancy, and multivariate dependencies are encoded within a single Hamiltonian constructed from mutual-information measures. By extending beyond quadratic interactions, the proposed approach is able to explicitly capture higher-order relationships between variables, which are typically inaccessible to standard feature selection methods.

A key advantage of this framework is that, once the cost function is defined, the method does not require model training. Instead, it relies purely on statistical dependencies in the data, making it inherently model-independent and broadly applicable across tasks. The only hyperparameter introduced in the workflow is the post-processing selection threshold, which is applied after quantum sampling to determine the final subset of features. 

From a hardware perspective, trapped-ion quantum processors such as IonQ Forte are particularly well suited for this problem due to their all-to-all connectivity, which naturally accommodates the long-range interactions induced by the HUBO formulation. This eliminates the need for complex embedding strategies and enables a direct implementation of higher-order feature selection Hamiltonians.

The experimental results demonstrate the effectiveness of the proposed approach. Across both the Gallstone and Spambase datasets, the method consistently reduces the dimensionality of the feature space (e.g., from 32 to 19 features and from 32 to 23 features) while maintaining, and in some cases improving, predictive performance. Moreover, the quantum-selected subsets outperform classical baselines such as SelectKBest based on mutual information and principal component analysis (PCA), highlighting the advantage of incorporating higher-order statistical structure in the selection process.

Looking forward, the proposed framework is expected to benefit significantly from advances in quantum hardware. As larger systems with increased qubit counts become available, it will be possible to tackle higher-dimensional feature selection problems, where classical approaches often struggle due to combinatorial complexity. In this context, next-generation trapped-ion devices, with improved scale and fidelity, together with the integration of BF-DCQO as an extension of the DCQO algorithm, represent a promising platform for practical quantum-enhanced feature selection in real-world machine learning pipelines.

\bibliographystyle{IEEEtran}
\bibliography{references}

\end{document}